\documentclass[12pt]{iopart}
\usepackage{color}
\usepackage{iopams}
\usepackage{epsfig}

\begin{document}

\title[The self-organizing impact of averaged payoffs on the evolution of cooperation]{
The self-organizing impact of averaged payoffs on the evolution of cooperation}

\author{Attila Szolnoki$^{1,*}$ and Matja\v z Perc$^{2,3,4,5}$}

\address{$1$ Institute of Technical Physics and Materials Science, Centre for Energy Research, Hungarian Academy of Sciences, P.O. Box 49, H-1525 Budapest, Hungary\\
$2$ Faculty of Natural Sciences and Mathematics, University of Maribor, Koroška 160, 2000 Maribor, Slovenia\\
$3$ Department of Medical Research, China Medical University Hospital, China Medical University, Taichung, Taiwan\\
$4$ Complexity Science Hub Vienna, Josefst\"adterstra{\ss}e 39, 1080 Vienna, Austria\\
$5$ Alma Mater Europaea, Slovenska ulica 17, 2000 Maribor, Slovenia\\
$*$ Author to whom any correspondence should be addressed.}

\ead{szolnoki.attila@energia.mta.hu}

\begin{abstract}
According to the fundamental principle of evolutionary game theory, the more successful strategy in a population should spread. Hence, during a strategy imitation process a player compares its payoff value to the payoff value held by a competing strategy. But this information is not always accurate. To avoid ambiguity a learner may therefore decide to collect a more reliable statistics by averaging the payoff values of its opponents in the neighborhood, and makes a decision afterwards. This simple alteration of the standard microscopic protocol significantly improves the cooperation level in a population. Furthermore, the positive impact can be strengthened by increasing the role of the environment and the size of the evaluation circle. The mechanism that explains this improvement is based on a self-organizing process which reveals the detrimental consequence of defector aggregation that remains partly hidden during face-to-face comparisons. Notably, the reported phenomenon is not limited to lattice populations but remains valid also for systems described by irregular interaction networks.
\end{abstract}

\maketitle

\section{Introduction}
\label{intro}

The presence or absence of cooperation has a huge consequence in various fields of life, therefore it has a paramount importance to identify which conditions help or which ones block the spreading of altruistic behavior in a complex population \cite{nowak_11,perc_pr17}. Interestingly, some universal mechanisms were identified in the last two decades which remain valid not only in microbiological systems, but also in human societies where interacting agents have significant cognitive skills to adjust their behavior for a higher individual income \cite{nowak_s06,sigmund_10,inaba_g19,liu_dn_pa19,sasidevan_srep16}.

Without exaggeration, hundreds of research papers were published by scientists with biology, economics, applied mathematics, or statistical physics background, in which they proposed different microscopic models to increase the general willingness of actors to cooperate with their partners \cite{szabo_pr07,richter_bs19,lin_zq_pa20,jiao_yh_csf20,cheng_f_pa19}. In some cases the desired evolutionary outcome is expected, for example when defection is punished or cooperation is awarded by individuals or by a governing institution \cite{gao_sp_pla20b,wang_sx_cnsns19,cong_r_srep17,wu_y_srep17,chen_xj_pcb18,liu_jz_csf18,liu_ls_mmmas19}. In these cases, however, the proper question is how to avoid the so-called second-order free-riding, when a cooperator player is reluctant to contribute and maintain the mentioned cooperation supporting institution or behavior \cite{helbing_ploscb10,perc_srep15,cheng_f_amc20,szolnoki_prsb15}. Intellectually it is more challenging to identify those mechanisms or conditions which do not support directly one of the competing strategies. More precisely, in the latter cases it is not obvious in advance why they result in a higher cooperation level. In these, so-called strategy-neutral alterations of the traditional models it is a common feature that the cooperator supporting effect emerges just as a secondary or indirect consequence of the fair and democratic rule. One of the very first and most celebrated example was to identify that heterogeneous population could be a cooperator supporting environment \cite{santos_prl05}. The heterogeneity may originate from an irregular interaction networks where some players have significantly more neighbors than for others hence they can collect higher payoff \cite{poncela_pre11,nagatani_jpsj20,yang_hx_epl18}. Diversity may also originate from different individual skills, like strategy teaching capacity or other social status, which could also result in similar effect \cite{szolnoki_epl07,perc_pre08,rong_zh_c19,pinheiro_rsos21,szolnoki_srep16}. The common feature behind these models is a kind of matching process in which players locally coordinate their strategies which reveal the advantage of cooperators. 
For completeness we note that coordination can be reached directly via a sort of conformity attitude \cite{zhang_lm_pa21,szolnoki_rsif15,yang_hx_csf17}, but it is not scope of our present work.
Similar impact can also be reached when player treat their neighbors differently, via weighted interaction graph, or support their neighbors in an unequal way \cite{meloni_rsos17,szolnoki_amc20,zhu_pc_epjb21,yang_hx_pa19,yang_hx_pa20,li_j_csf18}. Interestingly, an intervention into the microscopic dynamical process can also be helpful for cooperation. By introducing an inertia into the decision making or hindering too fast individual strategy change could also be beneficial for cooperation \cite{szolnoki_pre09,zhang_yl_pre11,szolnoki_csf20}. The related model studies pointed out that the mentioned dynamical change has asymmetric consequence on the invasion process of different strategies. The mentioned intervention does not relevantly modify the slow and balanced propagation of cooperator state resulting in smooth interfaces separating competing domains in a spatial system. On the other hand, the resulting dynamical change blocks significantly the rapid progress of defection state, which would lead to irregular interfaces and easy individual victory of defection otherwise. But we can also mention memory effects when accumulated success from past interactions can reveal that defection can only be successful for short term because neighbors who follow this behavior eliminate the potential prey of further exploitation \cite{fu_mj_pa19,danku_srep19,xia_cy_c20,xu_zj_c19}. 

In this work we consider an alternative ``strategy neutral'' modification of the traditional model where the positive consequence on the evolutionary outcome is not straightforward. In particular, we focus on the strategy imitation process when a learner player analyzes the payoff value of the model player who represents a tempting strategy. Evidently, the decision whether to adopt or not an alternative strategy is based on the information that learner player collects from the partner. But this info could be inaccurate \cite{szolnoki_njp14,szolnoki_epl15,hilbe_pnas18}. Principally we are not talking about a perception error, which can be handled by a noise parameter introduced in the strategy learning probability. Instead, we focus on the deceptive behavior of the model player. Such deception is rather frequent in animal kingdom and basically it serves to avoid conflicts or to gain mating advantage
\cite{searcy_06}. But of course, {\it Homo sapiens} is the best liar and we can easily give examples when someone's dress or lifestyle shows more than her/his proper success \cite{bond_jnb88}. This experience makes a general learner more careful who may try to collect additional information to evaluate an alternative strategy more accurately. In this way the learner's decision is not based solely on the success of a particular player but on a more reliable averaged statistics obtained from the neighborhood. Here the key question is how to weight the directly observed local and the average payoff values obtained from the learner's environment. Naturally, the size of the environment from which the learned collects information could also be a crucial detail.
To explore the possible consequences of extra information we study not just different sizes of perception environment of the learner player, but also check cases when the mentioned environment is not stable, but potential model players are chosen randomly from the population.

In the rest of this paper we propose a very simple model to explore how averaged payoff values change the learning process and reveal that it has a significant cooperator supporting consequence. We not just report this phenomenon, but also give a plausible explanation what is behind it. Furthermore, we also emphasize that the simple extension we propose results in a universally valid effect that could be observed in populations characterized by not only regular, but also irregular interaction graphs. But we first define our extended model, and then proceed with the results and a discussions of their implications for a more sophisticated and effective learning process.

\section{Evaluating the complete neighborhood}
\label{def}

We start from the traditional version of spatial prisoner's dilemma game model where players are distributed on a graph and interact with their neighbors. The players represent either cooperator ($C$) or defector ($D$) strategy, which strategies are distributed randomly in the initial stage. We first define our proposed model for a square grid, but the extension to other graphs is straightforward. For simplicity, but without jeopardizing the essence of the conflict of interests, we use the so-called weak prisoner's dilemma game parametrization where the only control parameter is $T$, temptation to defect, which characterizes a defector's income against a cooperator partner. The latter player gets nothing in the mentioned interaction, similarly to the case when two defector players meet. In the last case, when two cooperators meet, both collect $R=1$ payoff value. 

According to the standard simulation protocol, in an elementary step a randomly chosen player $x$, who has strategy $s_x$, plays the game with her neighbors and collects altogether $\Pi_x$ payoff from these interactions. Similarly, a neighboring player $y$, who has the opposite $s_y$ strategy, collects $\Pi_y$ payoff from the games played with the corresponding neighbors. In the usual strategy imitation rule the $\Pi_y-\Pi_x$ payoff difference has a crucial role on how likely player $x$ adopts the $s_y$ strategy of the model player $y$. This likelihood is defined by the well-known Fermi-function \cite{szabo_pre98}
\begin{equation}
\Gamma (\Pi_y-\Pi_x) = \frac{1}{ 1+\exp[-(\Pi_y-\Pi_x)/K]}\,,
\label{adopt}
\end{equation}
where $K$ denotes the noise factor which collects different sources of errors, like the possibility of a bad decision based on the available information. Our present work focuses on the reliability of information that can be collected from other partners. Of course, there are several ways how to deceive others for a particular reason. For instance, a player may try to show a different strategy to the neighborhood from the one she actually applies. But here we concentrate on the possibility that the payoff value we collect from a potential model actor is not accurate. Needless to say, such an ambiguity could be frustrating for the learner player because her decision about strategy change is based on this payoff value, as it is summarized by Eq.~\ref{adopt}.

To minimize the possible error of evaluating the competitor's payoff value, our learner player may want to collect alternative information about the potentially tempting strategy. More precisely, player $x$ makes a survey in the available neighborhood and checks the payoff values of all players who practice the alternative $s_y$ strategy. If player $x$ averages the related values then she has a more reliable information about the general success of the strategy she wants to adopt. Here we have two fundamental aspects to be contemplated. The first one is how strongly to consider the additional information collected from the neighborhood. This can be done in a way that we replace $\Pi_y$ in Equation~\ref{adopt} by a weighted $\Pi_w$ value which is the combination of the original $\Pi_y$ payoff value of model player $y$ and the $\Pi_{av}$ averaged value obtained from akin players from the neighborhood:
\begin{equation}
\Pi_w = q \Pi_{av} + (1-q) \Pi_y \,.
\label{weight}
\end{equation}
Here $q$ is the control parameter determining how strongly our learner player trusts on the alternative source of information about the success of tempting strategy. Accordingly, if $q=0$ then we get back the traditional spatial prisoner's dilemma game, while in the $q=1$ limit the adoption probability is based on the averaged value collected from the neighborhood exclusively. We must stress that the average is obtained by summing over not all payoff values detected in the neighborhood but are reduced to those values only which are achieved by similar strategies to the one represented by model player $y$. This way of averaging is in stark contrast to previously applied averaging methods \cite{wang_z_epl12,wang_z_srep13}, because our learner player does not want to explore the general wellness of the neighborhood, but focuses on the success of a specific strategy.
We also note that the averaging process is restricted to the payoff values of the alternative strategy in the neighborhood because the learner's goal to gain more accurate information for potential strategy update. This protocol is in stark contrast to general average process applied in mean-field calculations and in some previous spatial models \cite{yang_hx_cpl08,nagashima_amc19}. Evidently, to collect additional information from the neighborhood requires a high cognitive skill from a learner that was detected in previous human experiments \cite{boyd_pnas11,burton-chellew_prsb17,lamba_prsb14,guida_g21}.

\begin{figure}
\centering
\includegraphics[width=4.5cm]{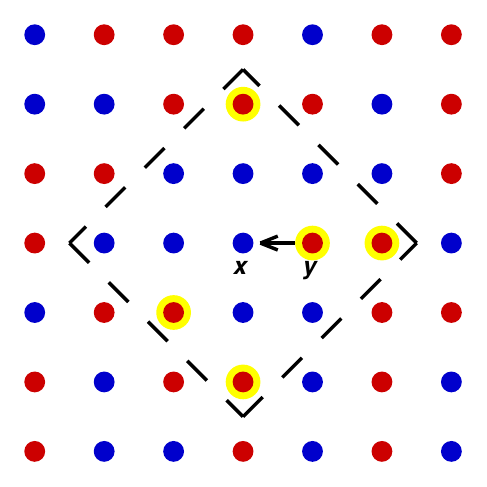}\\
\caption{As a learner, a cooperator (blue) player $x$ tries to imitate the strategy of the defector (red) neighbor player $y$. To calculate the imitation probability player $x$ considers not only the $\Pi_y$ payoff of player $y$, but also the $\Pi_{av}$ averaged payoff values of all other defector players who are within the evaluation circle. The latter players are marked by yellow background while the border of evaluation circle around player $x$ is marked by a dashed diamond. In this particular case $l_e = 2$ is applied, which means that all players whose distance from player $x$ are not larger than 2 may contribute to $\Pi_{av}$, hence providing a more accurate statistics about the general success of $s_y$ strategy.}\label{model}
\end{figure}

The other main ingredient of our model is to define the neighborhood of a learner player which is accessible to her to collect more accurate information. A natural way to assume that all players are checked who are within an $l_e$ steps from player $x$ hence they are within the evaluation circle. To clarify it better, we present a case in Fig.~\ref{model} where we surrounded by a dashed diamond shape line those players who are within the $l_e=2$ evaluation circle of the learner player $x$. Naturally, the value of $l_e$ can be increased from 1 toward higher values gradually and we can monitor how the information obtained from larger and larger set influences the evolution of cooperation. Importantly, as we already stressed, the $\Pi_{av}$ value is calculated from the values of those players who represent identical strategy to the one having by the model player $y$. In the above specified case they are marked by yellow background in our Figure.
 
%\begin{figure}[h!]

In our simulations we studied populations containing up to $N=160000$ players. According to the standard protocol in an elementary step a randomly chosen player has a chance to change her strategy by adopting the alternative strategy of a randomly chosen neighbor. By repeating this loop $N$ times we declare a natural unit of simulation, called 1 Monte Carlo (MC) step. In this work we applied maximum 50000 MC relaxation steps to reach the stationary states where the fraction of cooperators were measured and time averaged for another 100000 MC steps. 
The applied system size and the sufficiently long simulation time made us possible to obtain results which are independent of the applied system size, hence finite-size effects can be excluded. In this work we used $K=0.1$ noise level to allow comparison with results of traditional model, but we stress that our qualitative observations remain intact if we use other $K<2$ values of noise parameter. Of course, in the high $K$ parameter region the strategy imitation process becomes completely random and independent of the payoff difference.
Beside the mentioned square grid topology we also used random network interaction graph where the degree of nodes was $k=4$ unchanged. In this way we could check the consequence of using irregular topology without introducing additional effects originated from heterogeneity of players. Last, we mention that we also studied a case when the ``evaluation circle'' was not selected from players around the learner player, but we choose them randomly from the whole population. Nevertheless the details of this modified protocol will be given in the next section when we present its consequences by comparing with the results of the originally defined model.

\section{Results}

\begin{figure}[h!]
\centering
\includegraphics[width=8.5cm]{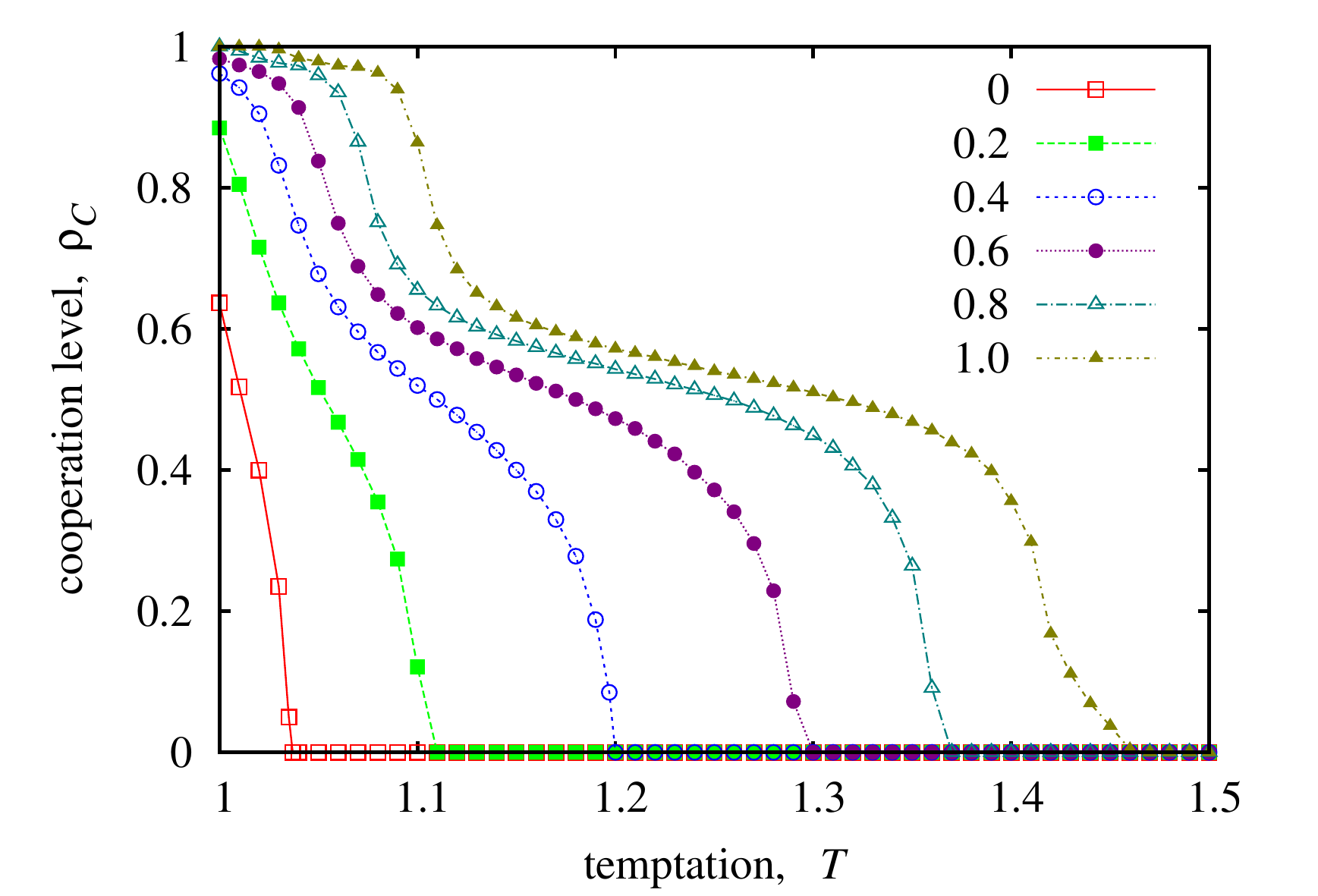}\\
\caption{Cooperation level in dependence of temptation value on a square lattice for different values of $q$ obtained at $l_e=2$. The values of $q$ are marked in the legend. The curves suggest that the cooperation level can be improved significantly if the learner players give greater credit to payoff information obtained from the neighborhood instead of trusting to the neighboring model player exclusively.}\label{sqr_q}
\end{figure}

Our first observations are summarized in Fig.~\ref{sqr_q} where we plotted the stationary cooperation level in dependence of temptation value for different values of the weight factor $q$. In the presented case we used $l_e=2$ evaluation circle, but qualitatively similar behavior can be found when the size of the neighborhood to collect extra information is different. As we noted here $q=0$ is equivalent with the traditional spatial model which suggests a $T_c=1.03576$ critical temptation value for the used $K=0.1$ noise level \cite{szabo_pre05}. But as we enlarge $q$, hence the learner players give a larger credit to the alternative information obtained from the neighborhood, the chance of cooperators to survive is improved significantly. Furthermore, when $q$ is close to 1, hence the additional information becomes dominant during the decision making about the strategy change then only $T>1.5$ temptation values can provide a full defection state. It is worth noting that $l_e$ is relatively small in the presented case, which practically means that typically not more a half dozen of other players are checked to gain a valuable extra information. Still, the improvement is remarkable.

\begin{figure}[h!]
\centering
\includegraphics[width=8.5cm]{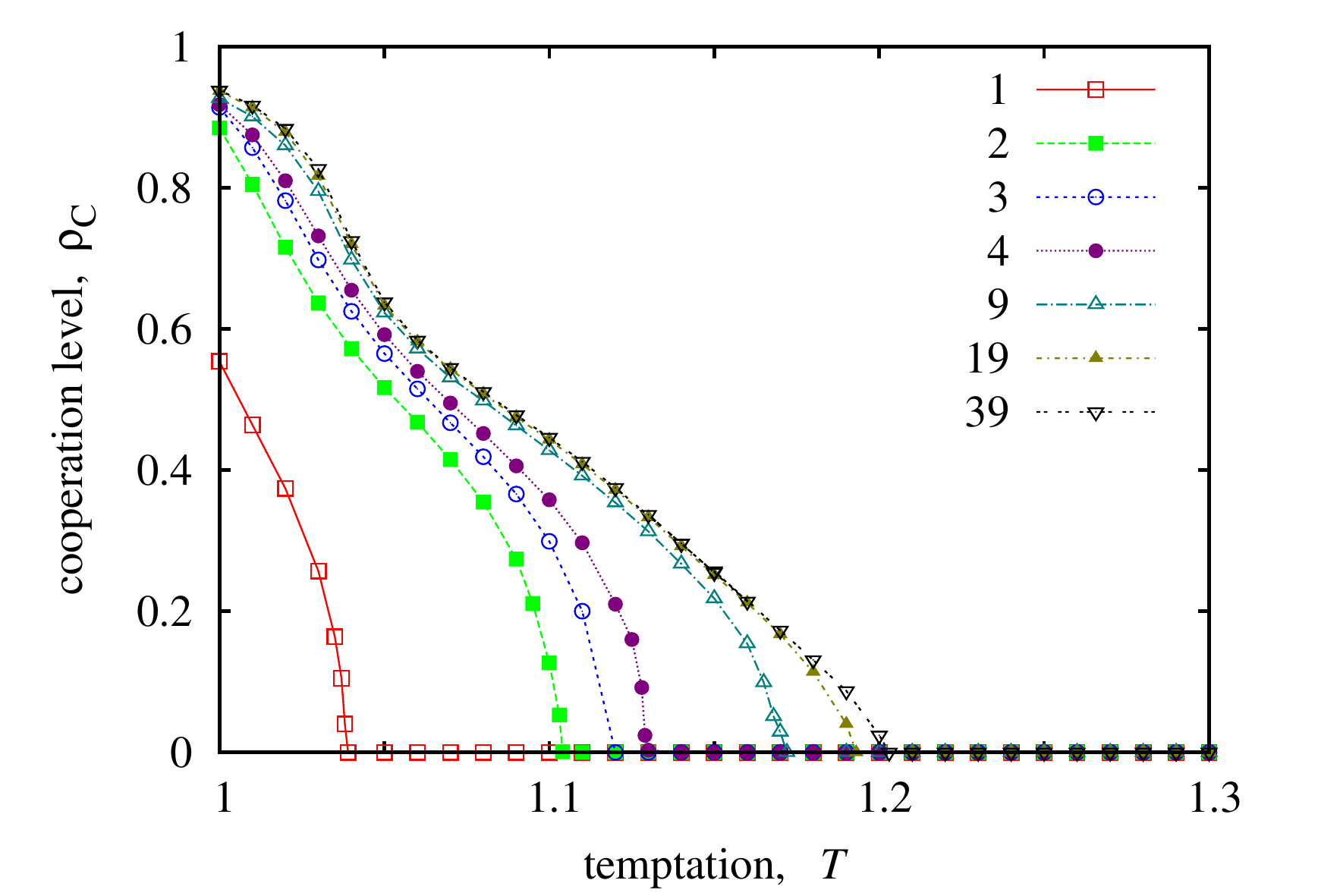}\\
\caption{Cooperation level in dependence of temptation value on a square lattice for different values of the radius of the evaluation circle obtained at $q=0.2$ weight factor. The values of $l_e$ are marked in the legend. The curves suggest that by enlarging the size of the neighborhood from the extra information is collected the cooperation level can be lifted even if the mentioned information has just a minor role on the decision making because of the small value of $q$.}\label{sqr_e}
\end{figure}

Next we illustrate how the size of the neighborhood, from which the extra information is gained, influences the cooperation level. A representative plot is shown in Fig.~\ref{sqr_e} where the weight factor is fixed at $q=0.2$. These curves highlight that the cooperation level can be improved if learners can collect information from a larger neighborhood. This effect, however, cannot be enlarged endlessly, because after a certain level this enhancement saturates. For example, collecting data from a 180-member set at $l_e=9$ gives almost equally good information than the neighborhood of beyond 3000 neighbors which is obtained for $l_e=39$. But the tendency is clear. One may note that the improvement of the critical temptation value characterizing the border of mixed state is not really large. But this change is the consequence of the relatively small $q$ weight factor which gives a modest credit to external information. We stress, however, that even at this $q$ the threshold $T_c$ can be doubled if the neighborhood size is large enough.

\begin{figure}
\centering
\includegraphics[width=13.5cm]{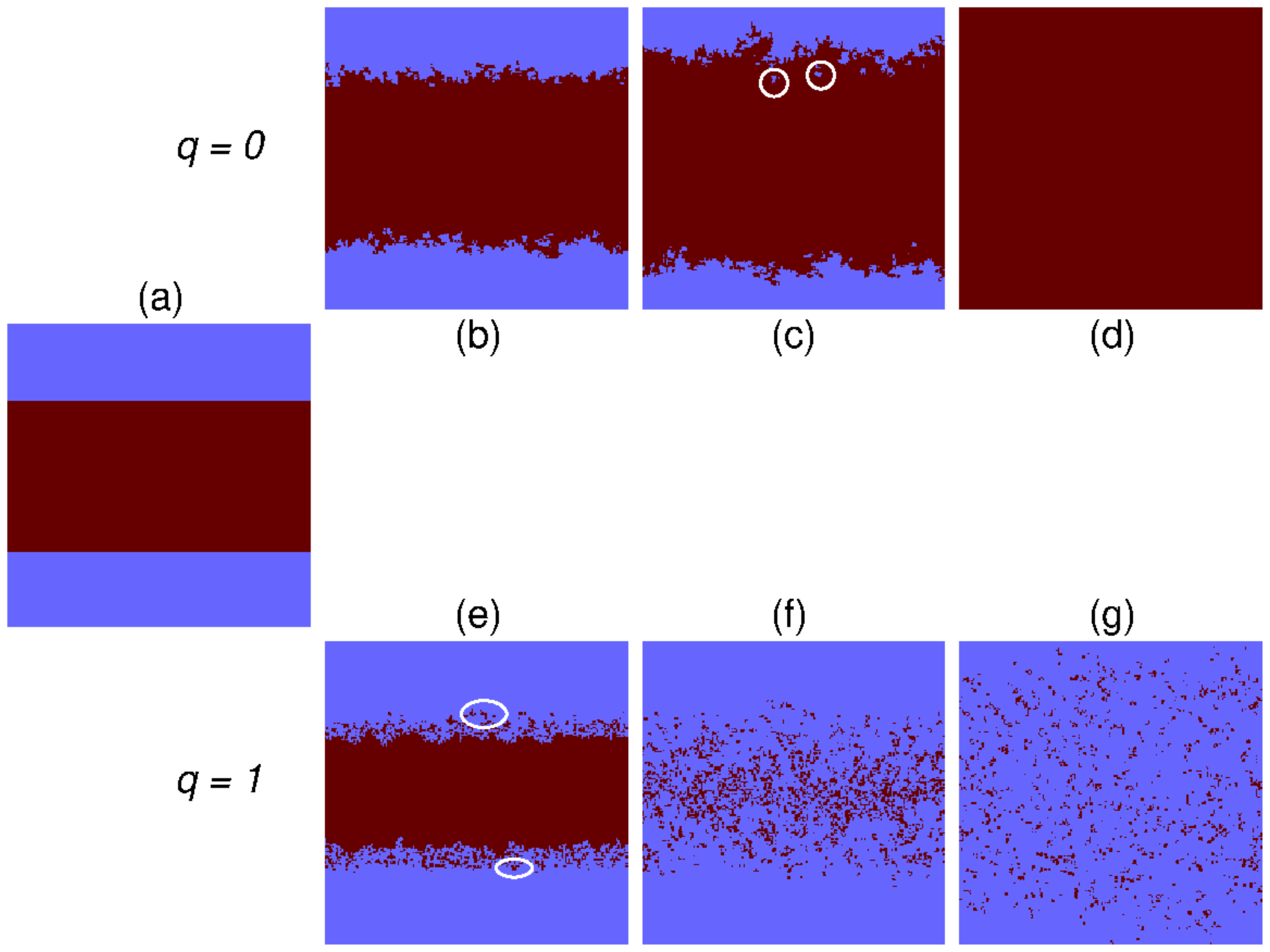}\\
\caption{Comparison of pattern formation for the traditional model (top row, panels~(b) to (d)) and the modified model (panels~(e) to (g)) where a learner considers a more accurate payoff value about the success of alternative strategy during the decision making. In both cases we applied the same $T=1.1$ temptation value and launched the simulation from an identical state, shown in panel~(a), where a red defector belt is surrounded by a blue cooperator domain. In the traditional case, which can be considered as $q=0$ case in the modified model, singular defectors can invade deep in the bulk of cooperator domain thanks to the high $T$ value which provides them high individual success. As a result, the interface separating the two main domains becomes irregular which makes the situations of cooperators more difficult. The moving front leaves behind it small cooperator islands, shown in panel~(c), but they cannot survive long and finally the system terminates into a full defector state. If $q=1$, shown in the bottom row, a defector's payoff in the front becomes less tempting, which  results in the shrink of the red belt. The propagating front leaves behind it small patches of defectors, shown in panel~(e), who can maintain a competitive payoff value, hence this phase eventually spreads in the whole system. In both cases we applied similar $L=200$ linear size.
}\label{single_neigh}
\end{figure}

To understand the cooperator supporting mechanism more deeply in the following we present a comparison of pattern formations obtained in the traditional and in the modified models. Figure~\ref{single_neigh} shows the significantly different evolutionary paths when we launch the simulations from the same initial state, shown in panel~(a), where a red defector island is surrounded by a blue cooperator domain. This setup, where different players meet along two domain walls, helps us to reveal the characteristic movement of propagation fronts more easily. For a proper comparison we applied the same $T=1.1$ temptation value for both cases. In the top row, containing panels~(b) to~(d), we show the evolution in the traditional model. Notably, this can be considered as a $q=0$ extreme case of the modified model. Here $q=0$ weight factor ensures that a learner $x$ player estimates the success of the alternative strategy based exclusively on the payoff value of her neighboring $y$ model player. As a consequence, shown in panel~(b) and (c), the original straight front line starts roughening because a neighboring defector can collect a high individual payoff value because of the relatively high value of temptation. We here note that the threshold temptation value is well below the presently applied $T$ value. The rough propagation front results in even more difficult circumstances for cooperators because it destroys their original phalanx and network reciprocity can hardly work anymore. Only just small islands of cooperators remain when the front passes. They are marked by white circles in panel~(c). However, they are unable to survive long because of the high $T$ value and the system eventually evolves to a full defector state, shown in panel~(d).

As a comparison, in bottom line we present the evolutionary path in the other extreme case, when the success of the alternative strategy is estimated from the information collected from the neighborhood. Despite of the fact that we applied relatively small $l_e=3$ radius, the mentioned trajectory is significantly different. Here the direction of the invasion is reversed by maintaining a not too noisy front line. Behind these lines, however, small fraction of defector players remain alive, as they are marked by white ellipses in panel~(d). The reverse direction of propagation informs us that a bulk of defector cannot provide a large average payoff to their members because there is no one to exploit. Furthermore, a pure cooperator domain provides robust average payoff value for a cooperating neighborhood, hence the adoption of cooperator strategy by a defector player becomes a frequent process in the initial stage. But if the density of defectors becomes low, as in the case marked by the ellipse, then they can collect competitive average payoff value again and form a stable coexistence with their rivals. Evidently, a pure cooperator neighborhood offers a high average of payoff, therefore the spreading of the mentioned mixed state is a slow process. For comparison, for $L=200$ linear system size the traditional evolution terminates into the full defector state typically within 300 MC steps, while at least 1000 MC steps needed to reach the stationary state in the modified case shown in panel~(g).

Based on the above described argument we can also understand why the $l_e$ value influences the stationary concentration of defector players. The larger the value of $l_e$ the smaller the faction of defectors who can survive permanently, as we observed in Fig.~\ref{sqr_e}. If their concentration exceeds a threshold value then their average payoff becomes less attractive, which provides a feedback mechanism to maintain a significant cooperation level in spite of relatively high temptation value. In this way the average information about the competing strategy maintains a self-organizing pattern of a mixed state where compelling cooperation level can be reached even for a high temptation value. This mechanism also explains our observations summarized in Fig.~\ref{sqr_q} because the effect becomes stronger as we give higher credit to the neighborhood via using larger $q$ weight factor.

One may ask what if the additional information is not collected from a local neighborhood, but originates from a random sampling where the target could be the whole population? In this modified model a learner player $x$ calculates the crucial $\Pi_{av}$ average payoff value by selecting $m$ other players randomly from the complete population. As previously, if the strategy of a selected $i$ player agrees to the strategy of the model player $y$ then we consider the related $\Pi_i$ payoff value of player $i$ when the mentioned average is calculated. Naturally, for a proper comparison with previously defined model extension, the value of $m$ should agree with the size of the neighborhood defined by the radius $l_e$. For example, if $l_e=1$ then $m$ should be 4, for $l_e=2$ the corresponding value is $m=12$, etc. The largest sampling set we used contains $m=3120$ members size is equivalent to a neighborhood around $x$ for $l_e=39$ radius. Importantly, the former $m=3120$ sub-population contains randomly chosen players from the whole population who are not necessarily neighbors to each others.

\begin{figure}[h!]
\centering
\includegraphics[width=8.5cm]{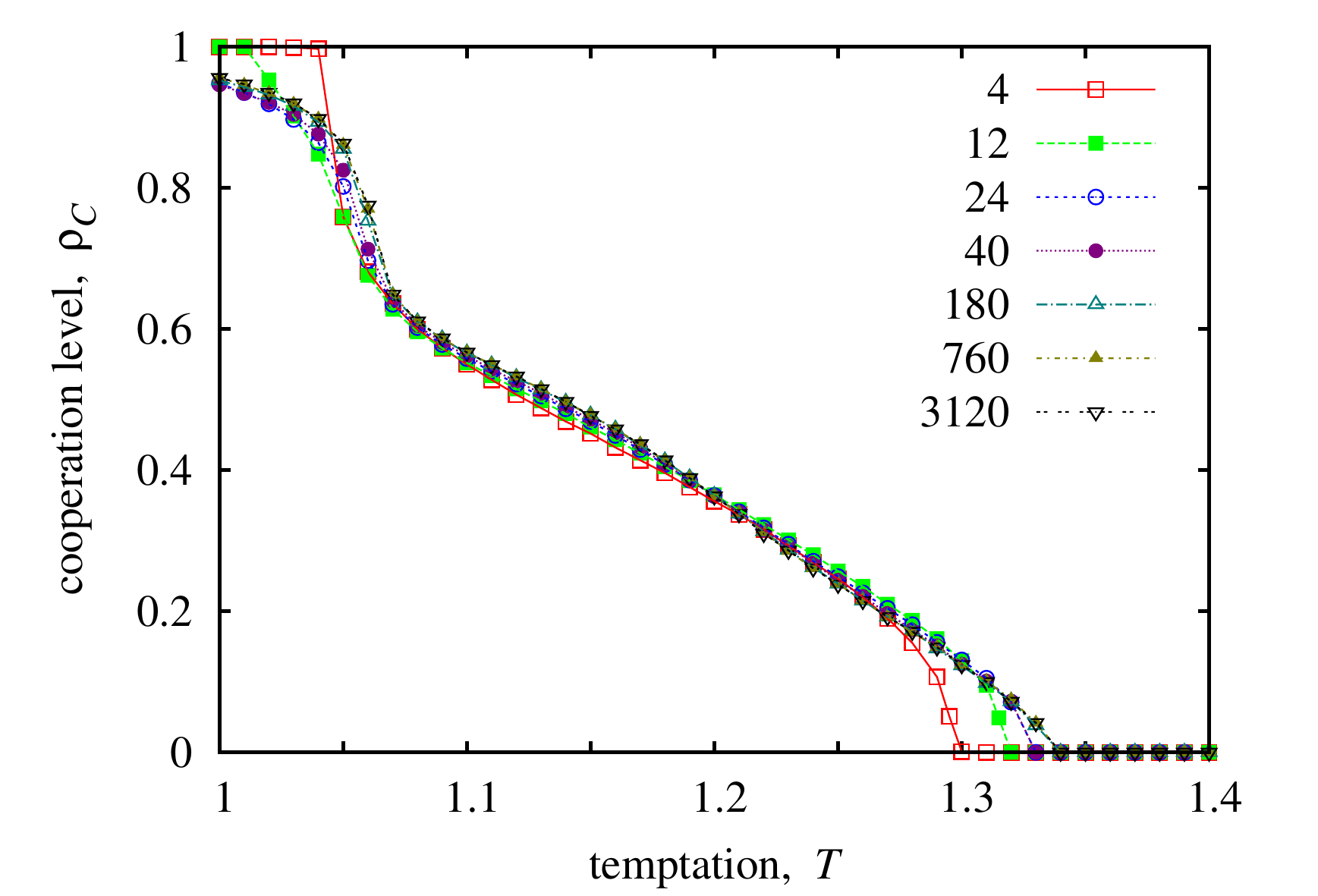}\\
\caption{Cooperation level in dependence of temptation value on a square lattice for different sizes of sample set obtained at $q=0.3$ weight factor. Players belonging to the sample set are chosen randomly for every learning process. Legend shows how many players are selected to gain information about the general success of the alternative strategy. The comparison of curves highlights that the size of sample set has no particular importance on the evolutionary outcome if we apply random sampling: the information gained from half dozen other players could be as valuable than the one gained from a much larger group. Here the most relevant parameter is the $q$ weight factor, which practically determines the critical temptation value until cooperators may survive. This value, however, is significantly larger than the one that can be reached when the information is collected from those players who are related to the learner player topologically.}\label{sqr_random_set}
\end{figure}

As previously, we still have two parameters, $q$ and $m$, but there roles are different from the one we previously observed for $q$ and $l_e$. A typical behavior is summarized in Fig.~\ref{sqr_random_set}. The first conspicuous feature is that size of the sampling set has no relevant role on the stationary value of cooperation level. Roughly speaking, it is enough to collect additional information from a small random sample because it gives no relevant advantage if a learner player bothers too much by checking too many players about the expected success of the alternative strategy. Our second observation is the general improvement of cooperation level comparing to the case when fixed and connected neighbors are used as a source of additional information. This fact can be seen easily if we check Fig.~\ref{sqr_q} where even a higher $q=0.4$ value is still unable to provide as high portion of cooperators as we see in Fig.~\ref{sqr_random_set} for random sampling.

\begin{figure}
\centering
\includegraphics[width=13.5cm]{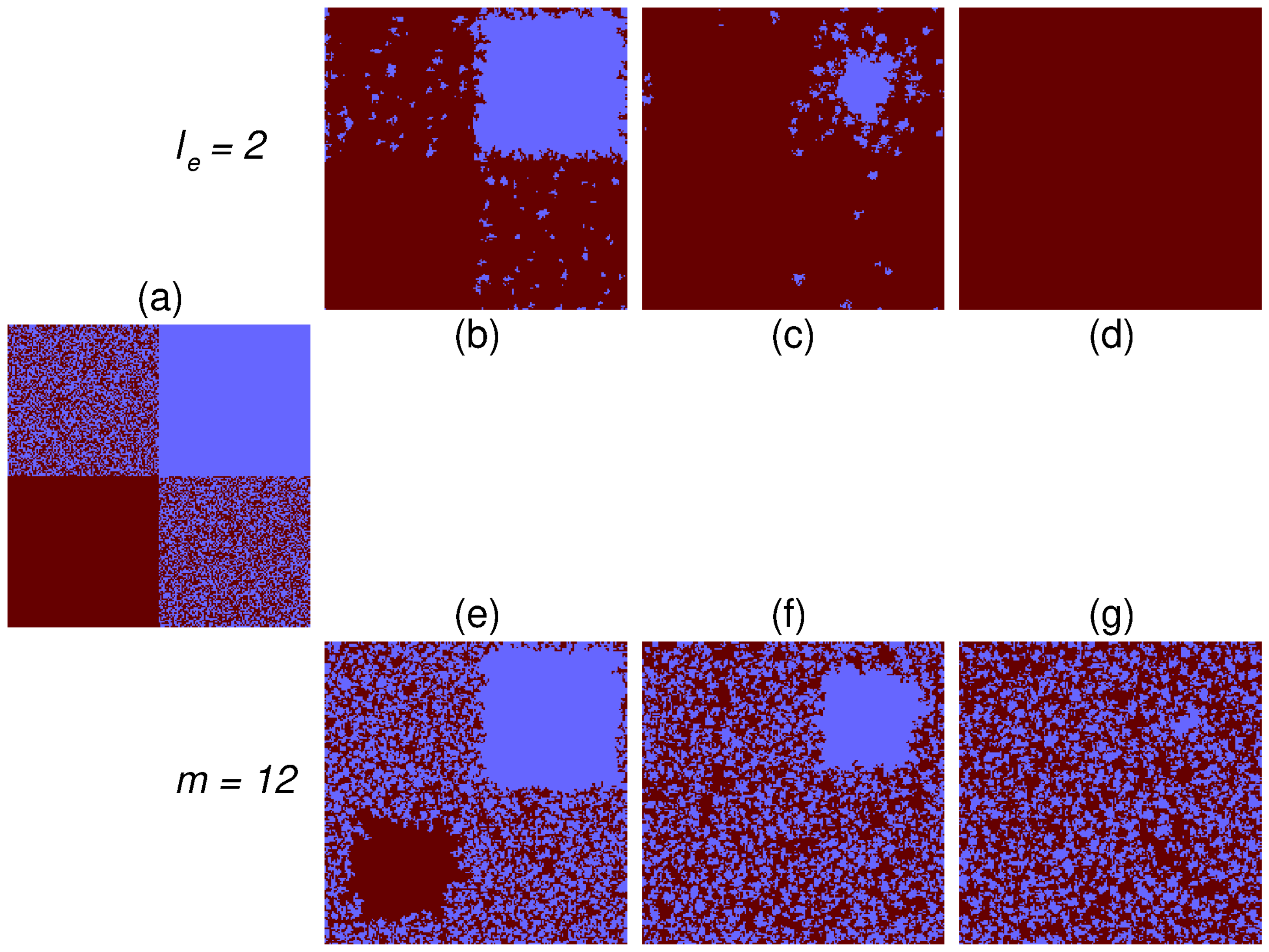}\\
\caption{
Comparison of pattern formation for the case when additional information is collected from the learner's neighborhood (top row, panels~(b) to (d)) and for the case when sampling players are chosen randomly (panels~(e) to (g)). For proper comparison we applied the same $T=1.3$ temptation value, $q=0.5$ weighting factor and used equally large sampling population as a source of additional information about the success of alternative strategy. The simulations were launched from identical starting state, shown in panel~(a). Initially we have a cooperator domain in the top-right corner and a homogeneous defector domain in the down-left corner. Their neighbor is a domain where strategies are distributed randomly. The hight temptation value cannot prevent the final victory of defectors in the fixed neighborhood sampling, while random sampling can provide a stable cooperation level shown in panel~(g). Further discussion can be found in the main text.
}\label{fixed_random}
\end{figure}

The above mentioned superiority of random sampling is valid for all related parameter pairs. Next we give some inspirations to understand its origin. To illustrate and understand the difference between random sampling and collecting data from a compact neighborhood in Fig.~\ref{fixed_random} we show how they drive the pattern formation at similar conditions. Importantly, we not simply apply equally strong temptation value and weighting factor, but also use equal size for the sampling population. Indeed, the latter has no decisive importance for random sampling, but it could be an essential factor for neighbor-based sampling, as it was illustrated in Fig.~\ref{sqr_e}. Accordingly, $l_e=2$ radius around the learner player is equivalent to check $m=12$ randomly chosen players. 
In contrast to Fig.~\ref{single_neigh} we here use an alternative common initial state, shown in panel~(a), where both a homogeneous cooperator and defector domains meet with a phase where strategies are mixed randomly. In the top row, where additional information is collected from the neighborhood, the fastest change can be observed in the mixed phase. This is a general phenomenon that can be seen even for spatially structured populations, because cooperators can only protect themselves if they are organized. In our case, despite of the relatively high $q=0.5$ weighting factor, network reciprocity alone is incapable to block the spreading of defection. It is because mixed environment can always give a decent payoff advantage for other defector players, too. At the same time a fully homogeneous and compact cooperator domain cannot really resist the invasion of defectors who are wrapped in a supporting cooperator shell. Interestingly, the homogeneous defector domain is not sensitive and cooperator player never enters into the down-left quadrant. At such a $q$ value the temptation is too high to replace defection by cooperation.

We, however, observe a strikingly different evolutionary trajectory when a learner collects information from randomly selected players. In this case the mix phase is table, albeit the actual ratio of defectors and cooperators is adjusted to the value of $T$ and $q$. On the other hand, the stability of homogeneous domains is proved to be the opposite we detected previously. Firstly, the shrink of the fully cooperator island is slower because in the average payoff of defectors may not be tempting: it can easily happen that we sample defector players from deep of the full-$D$ domain where they get nothing. But our argument is also valid for the opposite case. In the bottom row the homogeneous defector island becomes unstable and disappears very fast. In this case the strategy of cooperators standing at the front may become attractive because their average payoff value may be increased significantly by the contributions of their akin fellows who are sitting safe in the middle of a fully cooperator patch.

But we should stress that both dynamical process we discussed about the stability of homogeneous spots are just temporary because the distant information collected by random sampling drives the system eventually toward a uniform state where the fraction of $C$ and $D$ strategies is the same everywhere. In this stationary state, however, the previously mentioned self-organizing mechanism still works, which prevents defectors to grow too large homogeneous spot. Admittedly, this information gathering via random sampling also hinders cooperators to grow too large homogeneous domains because they cannot really utilize their high cooperation thanks to the smaller contributions to $\Pi_{av}$ from other $C$ players. Nevertheless, from cooperator's viewpoint the situation is fine because they can reach a decent fraction even at high temptation value if $q$ is large enough.

\begin{figure}[h!]
\centering
\includegraphics[width=7.5cm]{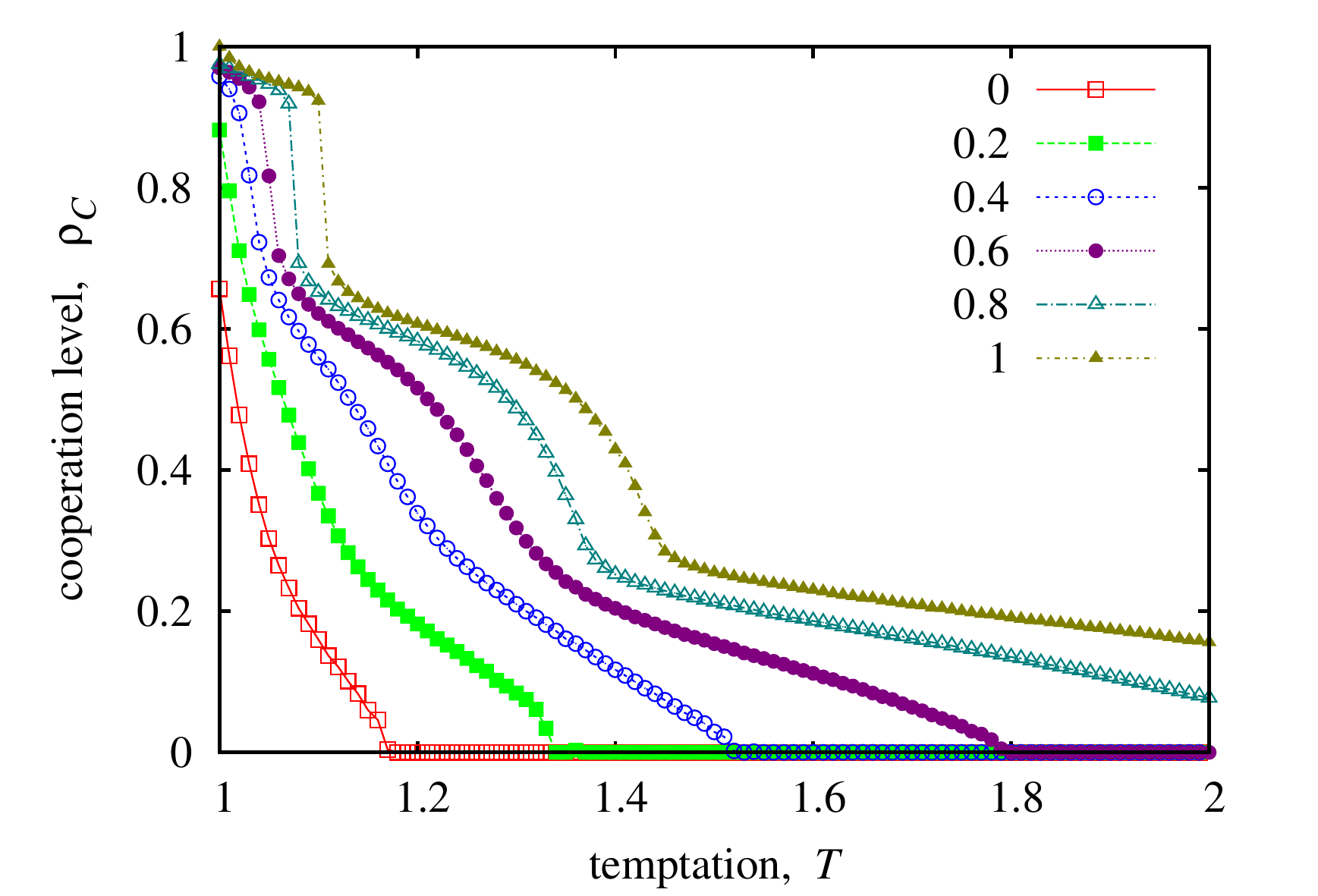}
\includegraphics[width=7.5cm]{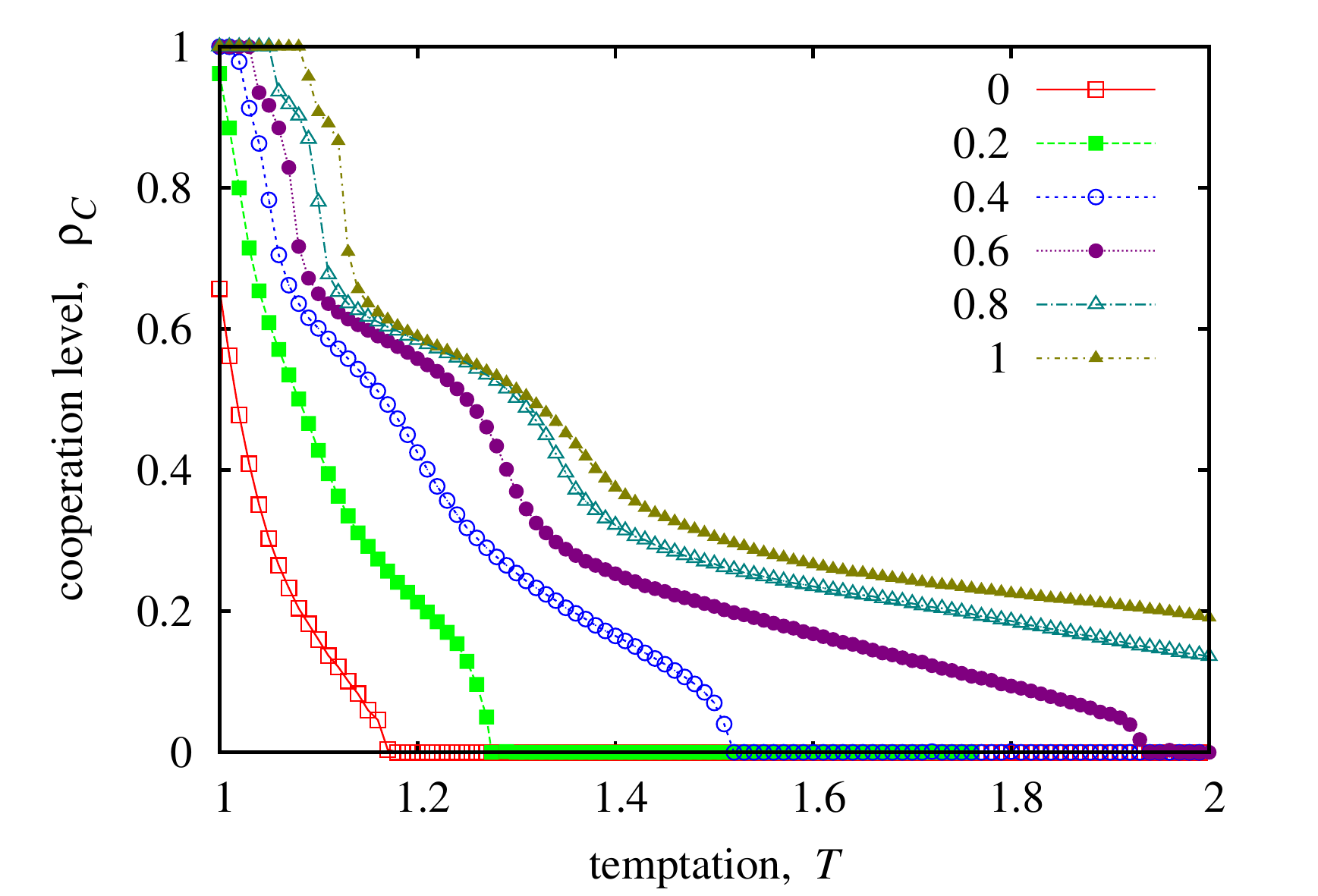}
\\
\caption{Cooperation level in dependence of temptation value on a random graph for different values of $q$ weight factor as indicated in the legend. For a proper comparison to previous results obtained for square grid, the degree distribution $k=4$ remained unchanged. On the left panel we show the case when the evaluation circle contains 12 nearest neighbor players of the actual learning player. On the right panel we show the results of a similar size of sampling set, but here these players are chosen randomly from the whole population. The comparison of the panels suggests that it has no significant importance whether the sampling set contains topologically related players or randomly chosen fellows. The only crucial factor is how large credit was given to the information collected from the external group when decision was made.}\label{rrg}
\end{figure}

Finally we briefly note that our observations about the positive consequence of considering additional information is not restricted to lattice-type populations, but remains valid when the interaction graph is not ordered. Having discussed the very positive consequence of random sampling, maybe this fact is not really surprising because our argument did not utilize the translation invariance of the interaction graph. But for completeness we also checked our results by using random topology where players have similar degree distribution as for square grid. Therefore we can check the consequence of randomness exclusively without bothering other effects due to the degree change of the topology.
The essence of our findings are summarized in Fig.~\ref{rrg} where we plot the results obtained for neighboring-based and random sampling based additional information gathering. 

In agreement to our earlier results, to give larger credit to averaged payoff values via increasing $q$ will improve the cooperation level. However, the clear consequence of random interaction topology is that this effect is really strong and cooperators may survive even at $T=2$. We stress that alone the random interaction topology would not be enough to produce such an improvement, because in the traditional model the critical temptation value remains close to $T=1$. Our other main observation is based on the comparison of the panels of Fig.~\ref{rrg} where similar curves are detected for neighboring-based and random sampling. This agreement suggests that the original randomness of the topology already serves as an information mixing tool. Therefore, in sharp contrast to the lattice-type topologies like square grid, in a randomized graph to collect extra information via random distance sampling has no additional value. But the positive consequence remains intact, and is more pronounced for irregular topologies.

\section{Conclusion}

Making a decision about which behavior to follow is a crucial act not only at personal but also on collective level. It is easy to see that the dominance of a hasty or careless adoption choice of players can drive the whole society toward an undesired destination. Therefore, huge intellectual efforts have been focused on this delicate task to find those methods which are in agreement with the fundamental Darwinian selection rule of the more successful strategy, but on the same time they help us to block the obvious advantage of defection. For example by recording and accumulating previous success of a strategy or by introducing an inertia and keeping a strategy more valuable if it survived long could be a cooperator supporting modification of the simplest ``imitating the more successful strategy'' protocol. But, of course, there are alternative methods and we refer the interested reader to related review papers.

In our present work we suggested a very simple modification of the traditional model where we considered the chance that a learner player is more careful and does not accept the information about the model player unreservedly. Instead, the former player tries to collect information about the success of the competing behavior from an alternative source. This could be the neighborhood of the learner player or could be randomly selected group of other players from the whole population. No matter which source is used, the population where members give higher credit to averaged information about the success of an alternative strategy can reach a higher cooperation level. The larger the weight of this additional information in the decision making the more significant improvement can be achieved.

The main mechanism which is responsible for this positive consequence is based on a self-organizing pattern formation of the spatial population. More precisely, to consider average information instead of accepting unconditionally the success of a particular case  prevents the condensation of defectors, hence maintains an acceptable cooperation level even at high temptation values. This procedure works not only in populations having lattice-type interaction graphs, but also for irregular topologies. 

It is worth noting that the observed cooperator supporting mechanism fits nicely to those where the introduced strategy-neutral rule has biased impact on the strategy invasion of competing strategies \cite{szolnoki_pre09}, hence they provide an alternative way to understand to original enigma, why cooperation may prevail among selfish agents. This research direction could be potentially promising for broader application of evolutionary game theory beyond human societies. In these systems participants may not necessarily have cognitive skills, like in microbiological populations, therefore the related theories should not rely on additional assumptions on moral issues, like reputation \cite{quan_j_pa21,yang_r_epjb20} or preliminary judgment about strategies which are the source of punishing or rewarding mechanisms in advanced populations \cite{yang_hx_epl20,wang_sx_pla21,gao_sp_pre20,takesue_epl18,gao_l_srep15}.

\ack
This research was supported by the Slovenian Research Agency (Grant Nos. P1-0403, J1-2457, and P5-0027)
\section*{References}

\providecommand{\newblock}{}


\begin{thebibliography}{10}
\expandafter\ifx\csname url\endcsname\relax
  \def\url#1{{\tt #1}}\fi
\expandafter\ifx\csname urlprefix\endcsname\relax\def\urlprefix{URL }\fi
\providecommand{\eprint}[2][]{\url{#2}}

\bibitem{nowak_11}
Nowak M~A and Highfield R 2011 {\em SuperCooperators: Altruism, Evolution, and
  Why We Need Each Other to Succeed\/} (New York: Free Press)

\bibitem{perc_pr17}
Perc M, Jordan J~J, Rand D~G, Wang Z, Boccaletti S and Szolnoki A 2017 {\em
  Phys. Rep.\/} {\bf 687} 1--51

\bibitem{nowak_s06}
Nowak M~A 2006 {\em Science\/} {\bf 314} 1560--1563

\bibitem{sigmund_10}
Sigmund K 2010 {\em The Calculus of Selfishness\/} (Princeton, NJ: Princeton
  University Press)

\bibitem{inaba_g19}
Inaba M and Takahashi N 2019 {\em Games\/} {\bf 10} 10

\bibitem{liu_dn_pa19}
Liu D, Huang C, Dai Q and Li H 2019 {\em Physica A\/} {\bf 520} 267--274

\bibitem{sasidevan_srep16}
Sasidevan V and Sinha S 2016 {\em Sci. Rep.\/} {\bf 6} 30831

\bibitem{szabo_pr07}
Szab{\'o} G and F{\'a}th G 2007 {\em Phys. Rep.\/} {\bf 446} 97--216

\bibitem{richter_bs19}
Richter H 2019 {\em BioSystems\/} {\bf 180} 88--100,

\bibitem{lin_zq_pa20}
Lin Z, Xu H and Fan S 2020 {\em Physica A\/} {\bf 553} 124665

\bibitem{jiao_yh_csf20}
Jiao Y, Chen T and Chen Q 2020 {\em Chaos, Solit. Fract.\/} {\bf 140} 110258

\bibitem{cheng_f_pa19}
Cheng F, Chen T and Chen Q 2019 {\em Physica A\/} {\bf 531} 121766

\bibitem{gao_sp_pla20b}
Gao S and Liang J 2020 {\em Phys. Lett. A\/} {\bf 384} 126723

\bibitem{wang_sx_cnsns19}
Wang S, Chen X and Szolnoki A 2019 {\em Commun. Nonlinear Sci. Numer.
  Simulat.\/} {\bf 79} 104914

\bibitem{cong_r_srep17}
Cong R, Zhao Q, Li K and Wang L 2017 {\em Sci. Rep.\/} {\bf 7} 14015

\bibitem{wu_y_srep17}
Wu Y, Chang S, Zhang Z and Deng Z 2017 {\em Sci. Rep.\/} {\bf 7} 41076

\bibitem{chen_xj_pcb18}
Chen X and Szolnoki A 2018 {\em PLoS Comput. Biol.\/} {\bf 14} e1006347

\bibitem{liu_jz_csf18}
Liu J, Meng H, Wang W, Li T and Yu Y 2018 {\em Chaos, Solit. and Fract.\/} {\bf
  109} 214--218

\bibitem{liu_ls_mmmas19}
Liu L, Chen X and Szolnoki A 2019 {\em Math. Models Methods Appl. Sci.\/} {\bf
  29} 2127--2149

\bibitem{helbing_ploscb10}
Helbing D, Szolnoki A, Perc M and Szab{\'o} G 2010 {\em PLoS Comput. Biol.\/}
  {\bf 6} e1000758

\bibitem{perc_srep15}
Perc M and Szolnoki A 2015 {\em Sci. Rep.\/} {\bf 5} 11027

\bibitem{cheng_f_amc20}
Cheng F, Chen T and Chen Q 2020 {\em Appl. Math. and Comput.\/} {\bf 378}
  125180

\bibitem{szolnoki_prsb15}
Szolnoki A and Perc M 2015 {\em Proc. R. Soc. B\/} {\bf 282} 20151975

\bibitem{santos_prl05}
Santos F~C and Pacheco J~M 2005 {\em Phys. Rev. Lett.\/} {\bf 95} 098104

\bibitem{poncela_pre11}
Poncela J, G{\'o}mez-Garde{\~n}es J and Moreno Y 2011 {\em Phys. Rev. E\/} {\bf
  83} 057101

\bibitem{nagatani_jpsj20}
Nagatani T and Ichinose G 2020 {\em J. Phys. Soc. Japan\/} {\bf 89} 064003

\bibitem{yang_hx_epl18}
Yang H~X and Yang J 2018 {\em EPL\/} {\bf 124} 60005

\bibitem{szolnoki_epl07}
Szolnoki A and Szab{\'o} G 2007 {\em EPL\/} {\bf 77} 30004

\bibitem{perc_pre08}
Perc M and Szolnoki A 2008 {\em Phys. Rev. E\/} {\bf 77} 011904

\bibitem{rong_zh_c19}
Rong Z, Wu Z~X, Li X, Holme P and Chen G 2019 {\em Chaos\/} {\bf 29} 103103

\bibitem{pinheiro_rsos21}
Pinheiro F, Pacheco J and Santos F 2021 {\em R. Soc. Open Sci.\/} {\bf 8}
  200910

\bibitem{szolnoki_srep16}
Szolnoki A and Perc M 2016 {\em Sci. Rep.\/} {\bf 6} 23633

\bibitem{zhang_lm_pa21}
Zhang L, Huang C, Li H, Dai Q and Yang J 2021 {\em Physica A\/} {\bf 561}
  125260

\bibitem{szolnoki_rsif15}
Szolnoki A and Perc M 2015 {\em J. R. Soc. Interface\/} {\bf 12} 20141299

\bibitem{yang_hx_csf17}
Yang H~X and Tian L 2017 {\em Chaos Soliton Fract.\/} {\bf 103} 159--162

\bibitem{meloni_rsos17}
Meloni S, Xia C~Y and Moreno Y 2017 {\em R. Soc. open sci.\/} {\bf 4} 170092

\bibitem{szolnoki_amc20}
Szolnoki A and Chen X 2020 {\em Appl. Math. Comput.\/} {\bf 385} 125430

\bibitem{zhu_pc_epjb21}
Zhu P, Hou X, Guo Y, Xu J and Liu J 2021 {\em Eur. Phys. J. B\/} {\bf 94} 58

\bibitem{yang_hx_pa19}
Yang H~X and Yang J 2019 {\em Physica A\/} {\bf 523} 886--893

\bibitem{yang_hx_pa20}
Yang H~X and Sun L 2020 {\em Physica A\/} {\bf 540} 123255

\bibitem{li_j_csf18}
Li J and Wang J 2018 {\em Chaos Soliton Fract.\/} {\bf 116} 1--7

\bibitem{szolnoki_pre09}
Szolnoki A, Perc M, Szab{\'o} G and Stark H~U 2009 {\em Phys. Rev. E\/} {\bf
  80} 021901

\bibitem{zhang_yl_pre11}
Zhang Y, Fu F, Xie G and Wang L 2011 {\em Phys. Rev. E\/} {\bf 84} 066103

\bibitem{szolnoki_csf20}
Szolnoki A and Chen X 2020 {\em Chaos Soliton. Fract.\/} {\bf 130} 109447

\bibitem{fu_mj_pa19}
Fu M, Guo W, Cheng L, Huang S and Chen D 2019 {\em Physica A\/} {\bf 525}
  1323--1329

\bibitem{danku_srep19}
Danku Z, Perc M and Szolnoki A 2019 {\em Sci. Rep.\/} {\bf 9} 262

\bibitem{xia_cy_c20}
Xia C, Gracia-L{\'a}zaro C and Moreno Y 2020 {\em Chaos\/} {\bf 30} 063122

\bibitem{xu_zj_c19}
Xu Z, Li R and Zhang L 2019 {\em Chaos\/} {\bf 29} 043128

\bibitem{szolnoki_njp14}
Szolnoki A and Perc M 2014 {\em New J. Phys.\/} {\bf 16} 113003

\bibitem{szolnoki_epl15}
Szolnoki A and Perc M 2015 {\em EPL\/} {\bf 110} 38003

\bibitem{hilbe_pnas18}
Hilbe C, Schmid L, Tkadlec J, Chatterjee K and Nowak M~A 2018 {\em Proc. Natl.
  Acad. Sci. USA\/} {\bf 115} 12241--12246

\bibitem{searcy_06}
Searcy W~A and Nowicki S 2006 {\em The Evolution of Animal Communication\/}
  (Princeton, US: Princeton University Press)

\bibitem{bond_jnb88}
Bond C~F and Robinson M 1988 {\em J. Nonverbal Behav.\/} {\bf 12} 295--307

\bibitem{szabo_pre98}
Szab{\'o} G and T{\H{o}}ke C 1998 {\em Phys. Rev. E\/} {\bf 58} 69--73

\bibitem{wang_z_epl12}
Wang Z, Szolnoki A and Perc M 2012 {\em EPL\/} {\bf 97} 48001

\bibitem{wang_z_srep13}
Wang Z, Szolnoki A and Perc M 2013 {\em Sci. Rep.\/} {\bf 3} 1183

\bibitem{yang_hx_cpl08}
Yang H~X, Wang B~H, Wang W~X and Rong Z~H 2008 {\em Chin. Phys. Lett.\/} {\bf
  25} 3504--3506

\bibitem{nagashima_amc19}
Nagashima K and Tanimoto J 2019 {\em Appl. Math. Comput.\/} {\bf 361} 661--669

\bibitem{boyd_pnas11}
Boyd R, Richerson P and Henrich J 2011 {\em Proc. Natl. Acad. Sci. USA\/} {\bf
  108} 10918--10925

\bibitem{burton-chellew_prsb17}
Burton-Chellew M~N, Mouden C~E and West S~A 2017 {\em Proc. R. Soc. B\/} {\bf
  284} 20170067

\bibitem{lamba_prsb14}
Lamba S 2014 {\em Proc. R. Soc. B\/} {\bf 281} 20140417

\bibitem{guida_g21}
Guida S~D, Han T~A, Kirchsteiger G, Lenaerts T and Zisis I 2021 {\em Games\/}
  {\bf 12} 25

\bibitem{szabo_pre05}
Szab{\'o} G, Vukov J and Szolnoki A 2005 {\em Phys. Rev. E\/} {\bf 72} 047107

\bibitem{quan_j_pa21}
Quan J, Tang C and Wang X 2021 {\em Physica A\/} {\bf 563} 125488

\bibitem{yang_r_epjb20}
Yang R, Chen T and Chen Q 2020 {\em Eur. Phys. J. B\/} {\bf 93} 94

\bibitem{yang_hx_epl20}
Yang H~X and Fu M~J 2020 {\em EPL\/} {\bf 132} 10007

\bibitem{wang_sx_pla21}
Wang S, Liu L and Chen X 2021 {\em Phys. Lett. A\/} {\bf 386} 126965

\bibitem{gao_sp_pre20}
Gao S, Du J and Liang J 2020 {\em Phys. Rev. E\/} {\bf 101} 062419

\bibitem{takesue_epl18}
Takesue H 2018 {\em EPL\/} {\bf 121} 48005

\bibitem{gao_l_srep15}
Gao L, Wang Z, Pansini R, Li Y~T and Wang R~W 2015 {\em Sci. Rep.\/} {\bf 5}
  17752

\end{thebibliography}
\end{document}